\documentstyle[twocolumn,aps,floats,amssymb,epsfig]{revtex}

\begin{document}

\newcommand{\bea}{\begin{eqnarray}}
\newcommand{\eea}{\end{eqnarray}}
 
\newcommand{\pone}{p_{t+1}}
\newcommand{\pmin}{p_{\min}}
\newcommand{\pt}{p_t}
\newcommand{\pmone}{p_{t-1}}
\newcommand{\ptk}{p_{t-k}}
\newcommand{\sumk}{\sum_{k=0}^{\tau -1}}
\newcommand{\pstar}{p^{\star}}

\newcommand{\qone}{q_{t+1}}
\newcommand{\qt}{q_t}
\newcommand{\qmone}{q_{t-1}}
\newcommand{\qtk}{q_{t-k}}
\newcommand{\epstar}{e^{-z \pstar}}
\newcommand{\epz}{e^{-z \pt}}
\newcommand{\eqz}{e^{-z \qt}}

\draft
\tolerance = 10000

\renewcommand{\topfraction}{0.9}
\renewcommand{\textfraction}{0.1}
\renewcommand{\floatpagefraction}{0.9}
\setlength{\tabcolsep}{4pt}

\def\d{{\rm d}}
\def\e{{\rm e}}
\def\i{{\rm i}}
\def\O{{\rm O}}
\def\half{\mbox{$\frac12$}}
\def\eref#1{(\protect\ref{#1})}
\def\etal{{\em et~al.}}
\def\Li{\mathop{\rm Li}}
\def\av#1{\left\langle#1\right\rangle}
\def\set#1{\left\lbrace#1\right\rbrace}
\def\stirling#1#2{\Bigl\lbrace{#1\atop#2}\Bigr\rbrace}

\twocolumn[\hsize\textwidth\columnwidth\hsize\csname @twocolumnfalse\endcsname

\title{A Simple Model of Epidemics with Pathogen Mutation}
\author{Michelle Girvan$^{1}$, Duncan S. Callaway$^{2}$, M.
E. J. Newman$^{3}$, and Steven H.
  Strogatz$^{2,4}$}
\address{$^1$Department of Physics, Cornell University,
Ithaca, NY 14853--2501}
\address{$^2$Department of Theoretical and Applied Mechanics,
Cornell University, Ithaca, NY 14853--1503}
\address{$^3$Santa Fe Institute, 1399 Hyde Park Road, Santa Fe, NM
87501}
\address{$^4$Center for Applied Mathematics, Cornell University,
Ithaca, NY 14853--3801}

\date{Received 18 May 2001}
\maketitle
\begin{abstract}

We study how the interplay between the memory immune response and
pathogen mutation affects epidemic dynamics in two related models.
The first explicitly models pathogen mutation and individual memory
immune responses, with contacted individuals becoming infected only if
they are exposed to strains that are significantly different from
other strains in their memory repertoire.  The second model is a
reduction of the first to a system of difference equations.  In this
case, individuals spend a fixed amount of time in a generalized immune
class.  In both models, we observe four fundamentally different types
of behavior, depending on parameters: (1) pathogen extinction due to
lack of contact between individuals, (2) endemic infection (3)
periodic epidemic outbreaks, and (4) one or more outbreaks followed by
extinction of the epidemic due to extremely low minima in the
oscillations.  We analyze both models to determine the location of
each transition.  Our main result is that pathogens in highly
connected populations must mutate rapidly in order to remain viable.

\end{abstract}

\vspace{1cm}
]

\section{Introduction}

The memory immune response enables humans and other animals to rapidly
clear, or even prevent altogether, infection by pathogens with which
they have previously been infected.  For example, we typically
contract chicken pox only once in our lifetime because of the
effectiveness of the memory immune response, and vaccines are designed
around the knowledge that our immune systems will more efficiently
fight foreign invaders if already exposed to something very
similar. Consequently, it is easy to imagine why some pathogens, such
as influenza, use a strategy of disguise to survive in a host
population.  In most cases, this disguise is facilitated by mutation:
pathogens permanently change their genetic content in order to alter
their appearance to the host immune system. With enough mutations, a
pathogen will ultimately be unrecognizable to the immune system of a
host that has previously been infected with one of its ancestors.  In
this paper, we study the dynamics of populations that lose immunity via
this route.

In epidemiological models, host populations are traditionally categorized
into three states: susceptible to infection ($S$), infected ($I$), and
removed or immune ($R$).  The succesion of states we will study is
depicted in the following diagram:

\begin{center}
\begin{picture}(100.0,40.0)
\put(0,25){\shortstack{$S$}}
\put(50,25){\shortstack{$I$}}
\put(100,25){\shortstack{$R$}}
\put(11,33){\shortstack[c]{{\scriptsize infection}}}
\put(61,33){\shortstack[c]{{\scriptsize recovery}}}
\put(23,3){\shortstack[c]{{\scriptsize loss of immunity}}}
\put(52.0,20.0){\oval(100.0,40.0)[b]}
\put(12,29){\vector(1,0){30}}
\put(62,29){\vector(1,0){30}}
\put(2,20){\vector(0,1){1}}
\end{picture}  
\end{center}

\noindent Models that describe such an epidemiological cycle are referred to as
$SIRS$ models.  While there is a vast literature covering models in which the
``loss of immunity" step is not considered (referred to as $SIR$ models; see
for example the classic texts by Bailey~\cite{bail75}, Anderson and
May~\cite{ande91} and the recent review by Hethcote~\cite{heth00}),
comparatively little work has been done to understand how the nature of the
$R\to S$ transition affects the dynamics of an epidemic.  

In principle, the transition depends on the strain to which one is
exposed (the challenge strain), in addition to one's previous history
of infection.  We thus begin our analysis with a computational
``bitstring model'' in which different pathogen strains are
represented by bitstrings that can mutate.  In this model, immunity
depends explicitly on the history of strains with which one has been
infected.  We find four fundamentally different types of behavior,
depending on parameters: (1) pathogen extinction due to lack of
contact between individuals, (2) endemic infection (steady state
infection), (3) periodic epidemic outbreaks (sustained oscillations),
and (4) one or more outbreaks followed by extinction of the epidemic
due to extremely low minima in the oscillations (``dynamic
extinction'').

We then develop a difference equation model in which the nature of
immunity is significantly simplified.  Instead of acquiring indefinite
immunity to a specific pathogen, individuals in this reduced model
spend a \emph{fixed} number of time steps in a generalized immune
class before being returned to the susceptible population. This
$SIR_1R_2\cdots R_{N}S$ model has been studied extensively by Cooke
\etal~\cite{cook77} and also by Longini~\cite{long80}, who used
stochastic methods to investigate the model with immunity lasting only
a single time step.  The same four qualitatively different dynamics
seen in the bitstring model are also observed for this model.  We
extend Cooke's work by deriving the location of the onset of
oscillatory dynamics in \emph{any} dimension (which is determined by
the number of time steps spent in the immune class).  Stability is
lost through a Hopf bifurcation, and changes in the model parameters
can increase the resulting limit cycle amplitude to the point that the
minimum becomes extremely small.  We establish a criterion for
``dynamic extinction'' (for which the minimum fraction infected in the
limit cycle oscillation is less than $1/N$, where $N$ is the
population size) and construct an asymptotic approximation for the
location of this extinction transition.

\section{Bitstring model}
\label{bitstring}

The immune response recognizes foreign molecules in a highly specific
way.  Individual immune cells or antibodies that recognize a protein
from one pathogen strain may be unable to recognize a similar protein
derived from another pathogen strain.  Because protein sequence and
structure is determined by the genetic content of an organism, immune
responses to a pathogen are in fact specific to the pathogen's genetic content.
We thus introduce a bitstring model in which pathogen strains are
represented by bitstrings, where the bitstring is regarded as an
abstract representation of a pathogen's genetic code.  Hosts are
immune to infection by pathogens that are highly similar to pathogens
with which they have been previously infected, where similarity
between two strains is measured by hamming distance \cite{hamming}.

The model consists of $N$ individuals who are either uninfected or
infected with a strain represented by one of the $2^\ell$ possible
bitstrings of fixed length $\ell$ \cite{footnote2}.  Individuals keep a record of all
the strains with which they have been infected, and we refer to these
histories as their memory repertoire.  Once per time step, each
infected individual exposes $z$ others by selecting individuals
uniformly at random from the entire population \cite{footnote3}. The
susceptibility of a contacted individual is determined by comparing
the bitstring of the challenge strain with the bitstrings of all
strains in the memory repertoire.  Specifically, an individual is
susceptible if $h_{\min}>h_{\text{thr}}$, where $h_{\text{thr}}$ is a
parameter, and $h_{\min}$ is the smallest hamming distance between the
challenge strain and any strain in the individual's memory repertoire.
With probability $\mu$ the challenge strain mutates by flipping one
randomly chosen bit; otherwise the strain remains unchanged.  In each
case considered here, we keep $\mu$ fixed at $0.1$ and vary the
threshold hamming distance $h_{\text{thr}}$; these two parameters are
inversely proportional.  Infection lasts a single time step, and
strain transmissibility is determined exclusively by individual immune
responses, i.e., in an entirely susceptible population no strain is
more fit than any other.

\begin{figure}[t]
\begin{center}
\epsfxsize=3in
\epsffile{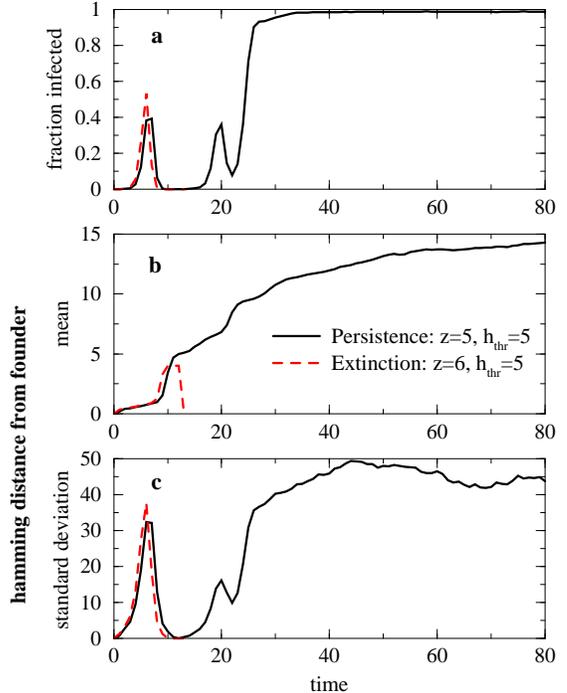}
\caption{The dynamics of infection in the bitstring model.  (a)
Fraction of population infected versus time for two different
threshold hamming distances.  For the smaller
value of $h_{\text{thr}}$, $p$ relaxes to a steady state value.
At the larger value, the disease does not persist. (b) The
hamming distance from the founder strain averaged over all
strains present. (c) The standard deviation of the distance
from the founder strain.  This is a measure of the diversity of
strains present.} 
\label{bitdynamics} 
\end{center} 
\end{figure}

In the first time step, one randomly chosen individual is
infected with the bitstring $000\cdots 000$ and all others have
never been infected.  From this initial condition we observe
three long term behaviors.  The first is trivial:  when $z<1$,
the size of the epidemic goes to zero since on average each
individual will expose fewer than one other individual.  The
remaining behaviors are (1) the fraction of the population that
is infected, $p$, approaches either a steady state nonzero
value or (2) after a brief outbreak, the epidemic dies out.
Figure~\ref{bitdynamics}a shows the difference in these two
types of behavior.  Interestingly, the transition from persistence
to extinction occurs as we increase $z$.  

\begin{figure}[t] 
\begin{center}
\epsfxsize=3in
\epsffile{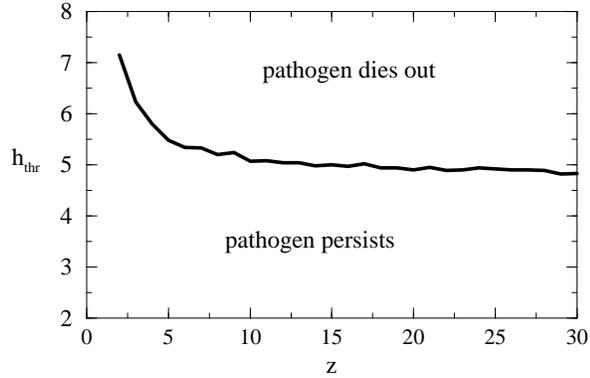}
\caption{Boundary between pathogen persistence and extinction in
the bitstring model.  Results were obtained by averaging the
greatest value of $h_{\text{thr}}$ at which the pathogen
persisted over 150 runs, for each value of $z$.} 
\label{bits_h_vs_Z}
\end{center}
\end{figure}

For a more complete characterization of this route to pathogen
extinction, we ran simulations at every integer--valued point in the
$z$--$h_{\text{thr}}$ plane with $0<h_{\text{thr}}<10$ and
$0<z<30$.  Figure 2 shows the result: below the curve, the
epidemic persists, and above the epidemic dies out.
Interestingly, $h_{\text{thr}}$ is a decreasing function of
$z$, meaning a high contact rate is actually detrimental to the
pathogen's ability to persist in a population.  We can explain
this result intuitively: the greater the value of $z$, the greater the
average number of previous infections any given host will have.  This
implies that every individual will have a larger memory repertoire, and
thus each strain's ability to infect a host will be reduced.  

Increasing $h_{\text{thr}}$ also causes extinction to occur.  This
happens because the number of strains to which an individual is
immune increases with $h_{\text{thr}}$, and thus reduces the likelihood
of infection.  Increasing $h_{\text{thr}}$ can also be thought of as
decreasing the pathogen mutation rate, since the memory immune response
will be more effective if the pathogen changes less rapidly.

When persistence occurs, the fraction of the population infected is
close to one (Figure~\ref{bitdynamics}a).  This is in contrast to
results from single strain $SIR$ models, which typically predict that
the fraction of infected individuals is proportional to (and less
than) $1-1/R_0$, where $R_0$ is the number of new infections that
would result from a single infection in a completely susceptible
population~\cite{ande91} -- this is $z$ in our case.  In
Figure~\ref{bitdynamics}a, the fraction infected at equilibrium is
approximately 0.99, certainly greater than $1-1/z=0.8$.  This
discrepancy occurs because there is more than one strain present in
the bitstring model, and individuals can be infected by two different
strains in two consecutive time steps.  Figures~\ref{bitdynamics}b--c
illustrate this effect: in the case when persistence occurs, the mean
hamming distance from the founder increases steadily in time, and
the diversity of strains present (measured by the standard deviation)
is high.  In contrast, when $z$ is increased and extinction occurs,
one can see that although the diversity is initially higher, prior to
extinction it is lower than in the persisting case.

\begin{figure}[t] 
\begin{center}
\epsfxsize=3in
\epsffile{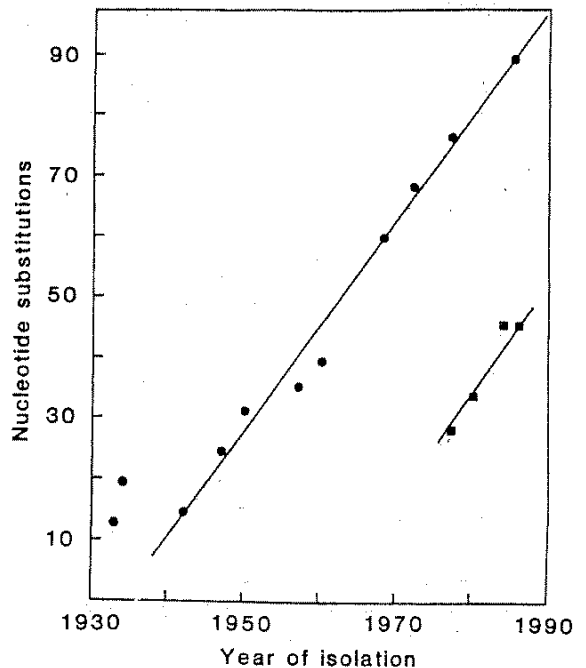}
\caption{Nucleotide substitutions in the nonstructural genes of
influenza A.  Figure reproduced with permision from Buonagurio
\etal~\protect\onlinecite{boun86}}
\label{boun_fig}
\end{center}
\end{figure}

In the absence of immunity, no single strain has a competetive
advantage over another in the bitstring model.  However, there is
evidence to suggest that in the case of influenza, there are selective
pressures that constrain base pair subsititions to a small part of the
entire genomic sequence space (i.e., with little diversity) and the
number of substitutions increases linearly in time~\cite{boun86}
(Fig.~\ref{boun_fig}).  This motivates us to consider a special form of
the model where bitstring mutation goes only in one direction, e.g.,
000...000 $\to$ 100...000 $\to$ 1100...000, etc.  Specifically, when
mutation occurs, rather than flipping a randomly chosen bit, the
leftmost zero in the bitstring sequence is changed to a 1.  All other
model mechanisms are as before.

\begin{figure}[t] 
\begin{center}
\epsfxsize=3in
\epsffile{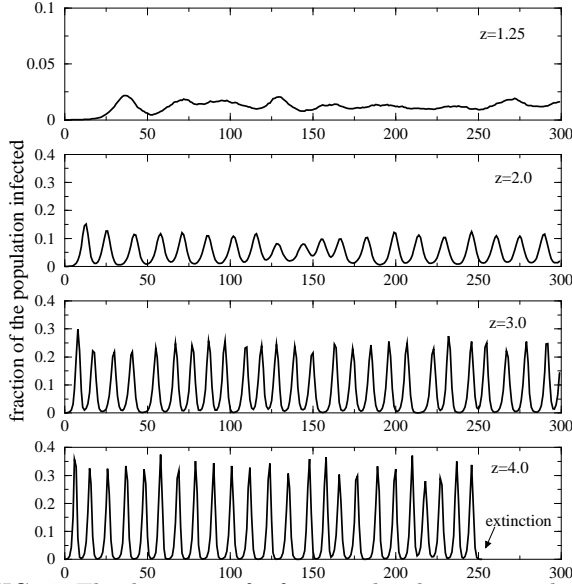}
\caption{The dynamics of infection when bitstring evolution is
constrained to go in one direction.  We have used $h_{\text{thr}}=4$.}
\label{lineardynamics} 
\end{center} 
\end{figure}

Figure \ref{lineardynamics} shows that under these assumptions, the
system's dynamics are quite different.  Most notably, as the
contact rate $z$ increases from $z=1.25$ to $z=2$, sustained
oscillations emerge from what appears to be steady state
behavior surrounded with stochastic noise.  As these
oscillations increase in amplitude, the minimum number of
infected individuals ultimately gets so low that the epidemic
dies out due to the population's finite size (at $z=4$).  

\begin{figure}[t] 
\begin{center}
\epsfxsize=3in
\epsffile{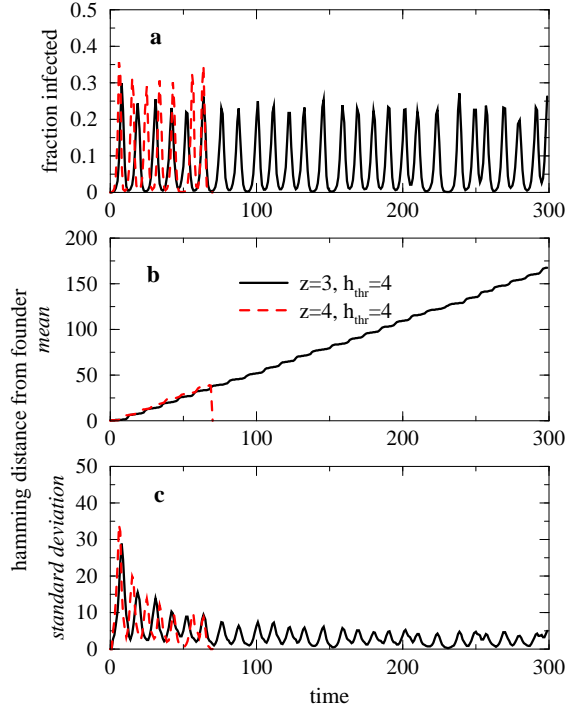}
\caption{The dynamics of infection in the bitstring model with
directed evolution.  (a)
Fraction of population infected versus time for two different
threshold hamming distances.  For the smaller value of
$h_{\text{thr}}$, $p$ converges to oscillatory behavior.  At
the larger value, the disease does not persist; the minimum
number of infected individuals gets too low. (b) The hamming
distance from the founder strain averaged over all strains
present.  The distance increases linearly with time, in
contrast to the results in Figure~\ref{bitdynamics}. (c) The
standard deviation of the distance from the founder strain.
This is a measure of the diversity of strains present.}
\label{hammingdynamics} 
\end{center} 
\end{figure}

Figure~\ref{hammingdynamics} depicts the dynamics in sequence
space for the last two cases from Fig.~\ref{lineardynamics}.  In
contrast to Figure~\ref{bitdynamics}, the hamming distance from the
founder strain increases linearly in time and the diversity is low.
These results mirror what has been observed for influenza, and motivate
us to understand the dynamics of this particular system in more detail.  

Under the assumption that pathogen evolution is constrained to be
linear in time, a further simplification of the system may be obtained
by assuming that individuals are immune for a fixed period of time
after infection.  In the following sections, we analyze a difference
equation model for this scenario to gain insight into the series of
transitions observed in Figure~\ref{lineardynamics}.

\section{Immunity Model}

\subsection{Model Derivation} \label{sec:map_derive}

In this section, we derive a system of difference equations to model
the average behavior of a closed population of susceptible, infective,
and immune individuals.  Upon infection, individuals spend one time step
in the infective class, and the subsequent $\tau-1$ time
steps in the immune class.  After these $\tau$ time steps,
individuals are returned to the susceptible pool.  The total
population size is $N$ and there are no births or deaths (i.e., $N$ is
constant).  

We define $p_{t+1}$ as the probability that an individual is
infected at time $t+1$:

\bea
p_{t+1}=x_t s_t,
\eea 

\noindent 
where $x_t$ is the probability that an individual is exposed at time
$t$, and $s_t$ is the probability of residing in the susceptible
class.  At each time step, the $Np_t$ infected individuals make $z$
random exposures.  The probability that an individual is not involved
in any such encounter is simply $(1-1/N)^{Np_tz}$, and so the
probability of exposure can be expressed as:

\bea
x_t=1-\left(1-\frac{1}{N} \right)^{Np_tz}.
\eea

The fraction of the population that is immune at any given time $t$,
can be determined by examining the fraction which has been infected in
any of the previous $\tau -1$ time steps, $\sum_{k=1}^{\tau -1}
p_{t-k}$.  The probability that an individual is susceptible is simply the
probability of being neither infected nor immune:

\bea
s_t=1-\sum_{k=0}^{\tau -1} p_{t-k}.
\eea
  
\noindent This gives

\bea
p_{t+1}=\left(1-\left(1-\frac{1}{N}\right)^{Np_tz}\right)\left(1-\sum_{k=0}^{\tau -1} p_{t-k}\right),
\eea

\noindent which simplifies to

\bea
\label{map}
\pone=(1-\epz)(1-\sumk \ptk)
\eea

\noindent in the large system limit, $N\to\infty$.  Thus, we have reduced the
immunity model to a $\tau$ dimensional map, or equivalently a system
of $\tau$ difference equations.  For large populations, the equations
are independent of $N$, and thus the only parameters are $\tau$ and
$z$.  This model was originally introduced by Cooke~\etal~\cite{cook77}.

\subsection{Dynamics}

\begin{figure}[t] 
\begin{center}
\epsfxsize=3in
\epsffile{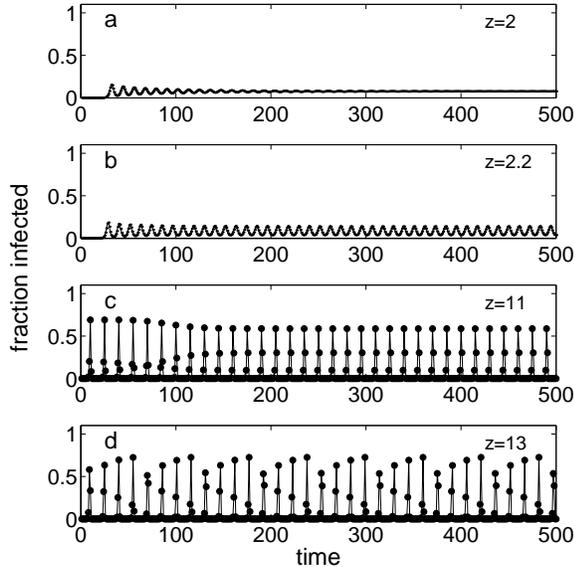}
\caption{Fraction of population infected versus time for $\tau$=6 and
various values of $z$.  (a) For $z=2$, $p_t$ relaxes to a fixed
point.  (b) For slightly larger values of $z$, $p_t$ exhibits small
amplitude quasiperiodic oscillations.  (c) At $z=11$, true period 15
behavior emerges.  (d) For $z=13$, $p_t$ exhibits
quasiperiodic oscillations that are heavily weighted by small values.}
\label{fig:map_pt}
\end{center}
\end{figure}

Numerical iteration of Eq.~\eref{map} from the initial conditions
$p_i=0, i=1 \dots \tau-1$ and $p_{\tau}=10^{-4}$ yields the same four
types of long term behavior observed in the bitstring model: approach
to the trivial equilibrium ($p_t\to 0$, not shown), approach to a
nonzero equilibrium (Fig.~\ref{fig:map_pt}a), sustained oscillations
that are generally quasiperiodic (Fig.~\ref{fig:map_pt}b--c), and dynamic
extinction (Fig.~\ref{fig:map_pt}d).  Dynamic extinction occurs only
due to numerical roundoff error, and is not an analytical feature of
Eq.~\eref{map}.  As we increase $z$ or $\tau$, $p_t$ eventually
becomes so small at the minimum of the oscillation that $(1-\epz)$ in
Eq.~\eref{map} numerically evaluates to zero.  Furthermore, oscillations
do not occur for $\tau \le 2$.

\begin{figure}[t] 
\begin{center}
\epsfxsize=3in
\epsffile{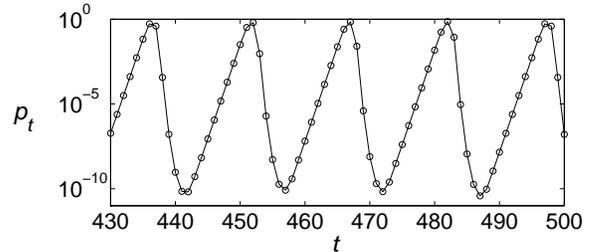}
\caption{Fraction of population infected versus time for
$\tau=6$ and $z=13$, plotted on a semilog scale.  The data set
used in the above plot is the same as the one used in
figure~\ref{fig:map_pt}d.}
\label{fig:exp_osc}
\end{center}
\end{figure}

For a fixed $\tau$, we observe that $p_t$ relaxes to a fixed point for
small enough $z$.  When $z<1$, $p_t$ approaches the trivial zero
solution.  As $z$ is increased through $z=1$, the non-zero equilibrium
becomes an attractor.  At some larger value of $z$, the fixed point
loses stability and small amplitude quasiperiodic oscillations appear
symmetrically centered around the former
equilibrium point.  As $z$ is further increased, the oscillations grow
in amplitude and the system spends a large fraction of its time with
only a small portion of individuals infected.  Careful examination of
the oscillations in this regime suggest they consist of two phases:
exponential growth followed by rapid decay (Fig.~\ref{fig:exp_osc}).  

For certain combinations of $\tau$ and $z$, the oscillations are truly
periodic rather than quasiperiodic.  We observe few patterns in the
location of these periodic solutions in the parameter space.  However,
we note that in the somewhat rare cases in which periodic behavior
emerges, the period is generally longer for larger values of $z$ and
$\tau$.  

\begin{figure}[t]
\begin{center}
\epsfxsize=3in
\epsffile{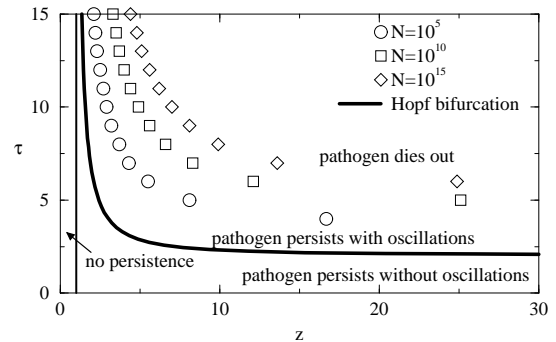}
\caption{The boundary between persistence and extinction for
different population sizes.  For each value of $\tau$, $z$ was increased until
the minimum expected fraction of infected individuals declined below $1/N$, for $N=10^5,
10^{10}$, and $10^{15}$.  The epidemic is considered extinct in the upper right
region of the figure because the minimum of the oscillation is below $1/N$. }
\label{mapdiesout}
\end{center}
\end{figure}

In the following sections we use linear stability analysis
to determine the location of the transitions from the trivial
equilibrium to the nonzero steady state and from the nonzero
steady state to oscillations.   The transition to dynamic extinction
is determined by deriving an approximate expression for the minimum
value of the oscillation and postulating that extinction occurs when
this value goes below $1/N$ where $N$ is any desired population
size. Figure~\ref{mapdiesout} illustrates the location of these
transitions in the $z-\tau$ plane.

\subsection{Stability Analysis}

The fixed points of the system satisfy 

\bea
\label{eq:pstar}
\pstar=(1-\epstar)(1-\tau \pstar).
\eea

\noindent Making the substition $q_t=p_t-\pstar$ and linearizing about
$\pstar$, we obtain

\bea
\qone&=& \Big( z \epstar (1- \tau \pstar) + \epstar -1 \Big) \qt
\nonumber \\
&&       + \Big( \epstar -1 \Big) \sumk \qtk.
\eea

\noindent Introducing the eigensolution $ \qt = q_{o} \lambda ^{t}$
yields a polynomial for the roots $\lambda_i$:

\bea
\label{eq:polylam1}
\lambda^{\tau} + \alpha \lambda^{\tau-1} +  \beta
\lambda^{\tau-2} +
           ... + \beta \lambda + \beta=0,
\eea

\noindent where $\alpha= 1 - \epstar( 1 +z -z \tau \pstar)$, and
$\beta=1-\epstar$.  

When the $\tau$ roots to this equation all lie within the unit
circle in the complex plane, the fixed point in question will be
stable.

\emph{{\bf Case i. \boldmath{$\pstar=0$}.}}  The first solution to
Eq.~\eref{eq:pstar} is

\bea
\pstar=0.
\eea

\noindent At this point, the eigenvalue equation simplifies to

\bea
\lambda^{\tau} -z\lambda^{\tau-1} = 0.
\eea

\noindent The first $\tau-1$ roots of this equation are $\lambda=0$.
The final solution solves $\lambda-z=0$, and thus the only
eigenvalue of interest for stability is

\bea
\lambda=z,
\eea

\noindent indicating that the trivial equilbrium is stable for all
$z<1$.  Not surprisingly, this agrees with our previous results from the
bitstring model in Section~\ref{bitstring}, since if $z<1$, fewer than
one new infection will result from each currently infected individual.

\emph{{\bf Case ii. \boldmath{$\pstar \ne0$}.}} The results in
Figure~\ref{fig:map_pt} suggest that when $z>1$, the nonzero solution to
Eq.~\eref{eq:pstar} becomes stable.  Indeed, Cooke~\etal~have shown for
all $z>1$ that this point is globally stable when $\tau=1$, and locally
stable when $\tau=2$.  They conjectured that when $\tau=3$, the fixed
point loses stability at $z=4.58$.  Here we will verify this result and
obtain the transition for all $\tau\ge 3$ by finding the location of the
bifurcation at which the quasiperiodic orbits emerge.  

As before, the onset of instability occurs when $|\lambda_i|=1$ for one
or more $\lambda_i$.  Therefore, to locate the transition, we substitute
$\lambda=e^{i \phi}$ into Eq.~\eref{eq:polylam1}.  This yields two new
equations, one for the real part and one for the imaginary part:

\bea
\label{eq:imag-real}
1- \epstar-(1-\tau\pstar)z\epstar  = \frac{\cos(\phi) -
\cos(\phi(\tau+1))}
                                        {\cos(\tau \phi) -1} \\
1-\epstar  =\frac{ 2 \sin(\phi)-\sin( 2 \phi)}
               {\sin ( \phi (\tau -1)) + \sin(\phi) - \sin( \tau
\phi)}
\eea

\noindent Combining these two equations with the expression for
$\pstar$ (Eq.~\eref{eq:pstar}), we have three equations for four
unknowns ($\pstar,\tau,z,$ and $\phi$).  These three equations
define the Hopf curve.

\begin{figure}[t]
\begin{center}
\epsfxsize=3in
\epsffile{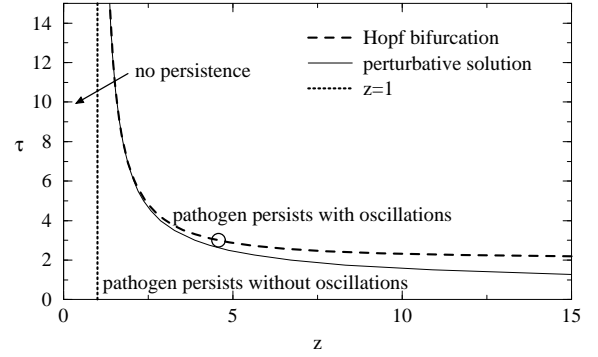}
\caption{ Hopf curve generated by numerically solving
equations~(\ref{eq:pstar}) and~(\ref{eq:imag-real}) simultaneously for
specified values of $\tau$.  The open circle marks the location of the
transition previously derived by
Cooke~\etal~\protect\onlinecite{cook77}.  The solid line is our
asymptotic solution, taken to $O(1/\tau^3)$.}
\label{hopf}
\end{center}
\end{figure}

Figure~\ref{hopf} shows the numerically generated solution for the
Hopf curve in the $z$--$\tau$ plane.  The dependence on parameters is
clearly similar to what we observed in the biststring model as that
system switched from persistence to extinction.  For $z<1$,
$\pstar=0$ is the only attractor.  When $z>1$ below the Hopf curve,
the stable fixed point is given by the non--zero solution to
Eq.~\eref{eq:pstar}. 

For large $\tau$, we can write a perturbative solution for the Hopf
curve.  Defining a new variable $\epsilon=\frac{1}{\tau}$, we can
express $\phi,z,$ and $\pstar$ as power series in $\epsilon$:

\bea
z&=&1+\sum_{i=1}^{\infty}a_i \epsilon^i, \\
\pstar&=&\sum_{i=2}^{\infty}b_i \epsilon^i, \\
\phi&=&\sum_{i=1}^{\infty}c_i \epsilon^i.
\eea

Solving perturbatively for the $a_i$, $b_i$, and $c_i$ gives

\bea
z&=&1 + \frac{\pi^2}{2} \epsilon + \frac{3 \pi^2}{4} \epsilon^2
  +\frac{5}{96}(12 \pi^2 +\pi^4)\epsilon^3+O(\epsilon^4),\\
\pstar&=&\frac{\pi^2}{2}\epsilon^2+\frac{1}{4}\pi^2 (2-\pi^2) \epsilon^3
  +O(\epsilon^4),\\
\phi&=& \pi \epsilon +\frac{\pi}{2} \epsilon^2 + \frac{\pi}{4} \epsilon^3
  + O(\epsilon^4).
\label{phieqn}
\eea

\noindent Figure~\ref{hopf} shows that for large $\tau$, the perturbative
solution for the bifurcation curve agrees well with the numerical
solution.

The variable $\phi$ may be thought of as the rotation
number~\cite{guck83} of the solution to the linearized equations.  In
cases where a periodic orbit emerges at the bifurcation, $2\pi/\phi$
will be the period of the orbit, and in cases where the orbit is
quasiperiodic, the approximate pattern will repeat itself on average
every $2\pi/\phi$ time steps.

Eq.~\eref{phieqn} indicates that $\phi$ is a decreasing function of
$\tau=1/\epsilon$, meaning that the period of the epidemic
oscillations will increase with $\tau$.  In other words, we expect
oscillations to occur less frequently as the duration of immunity
increases.

\subsection{Relaxation Oscillations and the Route to Extinction}

As pointed out earlier, Figure~\ref{fig:exp_osc}
suggests that as $z$ and $\tau$ become large, the system's dynamics
can be characterized by exponential growth followed by rapid decay,
with short transitional phases between the two regimes.  In this
section, we explore the relaxation oscillations by deriving equations
that approximate the system's behavior in the various phases.

\begin{figure}[t]
\begin{center}
\epsfxsize=3in
\epsffile{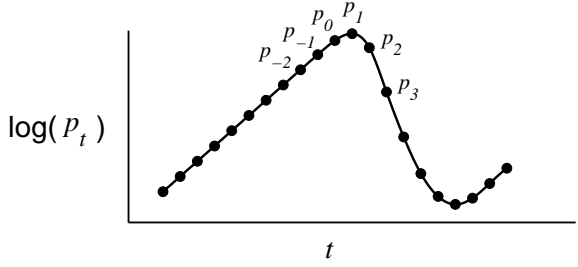}
\caption{Generic form of a relaxation oscillation for large $z$ and
$\tau$.  The oscillation is characterized by exponential growth 
followed by rapid decay.}
\label{relax_single}
\end{center}
\end{figure}

We begin by focusing on a single oscillation as pictured in
Figure~\ref{relax_single}.  As $p_t$ grows exponentially toward its
maximum, the fraction of individuals infected at the current time and
the previous $\tau-1$ time steps is small enough that, to a first order
approximation, Eq.~\eref{eq:pstar} can be rewritten as

\bea
\pone \approx zp_t \qquad \textrm{(in the growth phase)}.
\label{growth_approx}
\eea

\noindent The behavior persists until $z p_t$ reaches order 1.  We make
the assumption that Eq.~\eref{growth_approx} holds until the point
$p_0$ and check for consistency after subsequent calculations.  For
times less than $t=0$, we can write $p_{-t} = p_0/z^t$.

Next, we iterate the $\tau$--dimensional map using the unkown form of
$p_0$ to find $\tilde{p}_{\min} ( p_0)$.  The tilde notation
indicates that $\tilde{p}_{\min}$ is a local minimum of the
oscillations.  To find the global minimum, we must find the value of
$p_0$ that minimizes $\tilde{p}_{\min}$.  In what follows, we derive an
approximation for $p_{\min}$ in the large $z$ limit.  This calculation
requires that the points near the transition phase be handled
individually before a general expression for the decay behavior may be
obtained.  We determine $p_1$, $p_2$, and $p_3$ explicitly and then
derive the general form for $\pone$ for $t \ge 3$.

A reasonable approximation for $p_1=x_0 s_0$ is obtained by including only
the first term that appears in the sum in $s_0$:

\bea
p_1 \approx \left( 1 - \exp(-z p_0) \right) (1 - p_0)
\eea

Next, $p_2$ can be determined by first considering the probability of
exposure at time $t=1$.  Since the fraction of individuals infected is
order one at the maximum point of the oscillation, the
probability of exposure at time $t=1$ goes to one in the limit of
large $z$.  This implies that virtually
all individuals who are susceptible at time $t=1$ will become infected
at time $t=2$: $p_2\approx s_1$.

In order to produce a simple expression for $p_2$, it useful to note
that the fraction of individuals which reside in the susceptible class
at any point in time can be written in terms of the same fraction at
the previous time step:

\bea
s_t &=& \exp (-z p_{t-1}) s_{t-1} + p_{t-\tau}.
\eea

\noindent Using this, we write $p_2$ in a convenient form:
\bea
p_2 \approx s_1 \approx \exp( -z p_0) ( 1 -p_0) + \frac{p_0}{z^{\tau -1}}.
\label{p_2}
\eea

To find $p_3$, the final point in the transition region, we note that
$s_2 \approx p_{2-\tau} \approx \frac{p_0}{z^{\tau -2}}$, since $z p_1 \gg 1$.
Assuming $z p_2$ is small, we obtain:

\bea
p_3 \approx z p_2 s_2 = \frac{p_0}{z^{\tau -3}} p_2.
\label{p_3}
\eea

The general behavior in the decay regime can be derived by examining
the fraction of individuals susceptible for $ 3 \le t \le \tau$.
In this region, $s_{t-1}$ is small compared to $p_{t-\tau}$, yielding:

\bea
s_t &\approx& p_{t-\tau} \approx  \frac{p_0}{z^{\tau -t}} 
\qquad \textrm{(for $3<t<\tau$)}. 
\eea

Since $z p_t$ is small in the decay phase, the probability of exposure
depends linearly on the fraction of individuals infected: $x_t \approx
z p_t$.  Thus, we see that for $t \ge 3$, we can again replace the
original set of $\tau$ difference equations by a single equation:

\bea
\pone = z^{t -\tau +1} p_0 p_t .\qquad \textrm{(for $3<t<\tau$)}.
\label{decay-approx}
\eea

The minimum of $p_t$ occurs when $z^{t -\tau +1} p_0$ becomes greater
than 1.  If we assume that $p_0$ lies between $\frac{1}{z}$ and 1, as
is consistent with our earlier assumption that
Eq.~\eref{growth_approx} holds until $t=0$, then the minimum must
occur at $t=\tau$.  Consequently, we can express $p_t$ in terms of
$p_3$ for times greater than $t=3$.  Combining
Eqs.~\eref{decay-approx} and \eref{p_2}, we obtain:

\bea
p_{\tau} &=& z^{-\frac{1}{2}(\tau^2 - 5 \tau + 6)} p_0^{\tau -2} p_2
\label{p_tau}
\eea

Finding the minimum value of $p_{\tau}$ requires solving for the roots
of the equation $\d p_{\tau}/\d p_0 = 0$, which yields

\bea
(\tau-2) \left( \frac{p_0 \exp(z p_0)}{z^{\tau-1}} + 1 - p_0 \right)
&&
\nonumber \\
+ p \left(  \frac{\exp(z p_0)}{z^{\tau-1}} -1 -z(1-p_0) \right) &=&0
\label{roots}
\eea

\noindent We try a solution of the form

\bea
p_0 = \frac{\log (f)}{z}. 
\eea

\noindent Substituting this into Eq.~\eref{roots} we obtain an asymptotic expression
for $f$, which is valid to leading order as $z\to\infty$: $f \approx
z^{\tau}/(\tau -1)$.  This gives:

\bea
p_0 &\approx& \frac{\tau \log(z)}{z},
\label{p_0}
\eea

\noindent which is consistent with our assumption that $p_0$ must lie between $1/z$ and
1.  



Finally, to obtain $\pmin$, we insert (\ref{p_0}) into our
expressions for $p_2$ and $p_{\tau}$ (Eqs.~\eref{p_2} and \eref{p_tau}):

\bea
\pmin = z^{-\frac{1}{2}(\tau^2 -  \tau + 2)} 
\left(\tau \log(z)\right)^{\tau -1}.
\eea

In the derivation of the difference equations in Section
\ref{sec:map_derive}, we assumed an infinite population size.  Under
this assumption, the fraction of immune individuals can be arbitrarily
close to 1 without driving the disease to extinction.  If, however,
we assume the population size is finite, at some point the minimum
value of $p_t$ will be less than the fraction of the population
equivalent to one individual, that is $p_{\min} < 1/N$.  Replacing
$\pmin$ by $1/N$, we have an equation for the curve that
separates the region of disease persistence from the region of
extinction:

\bea
\frac{1}{N} = z^{-\frac{1}{2}(\tau^2 -  \tau + 2)} 
\left(\tau \log(z)\right)^{\tau -1}.
\label{extinct_predict}
\eea

\begin{figure}[t]
\begin{center}
\epsfxsize=3in
\epsffile{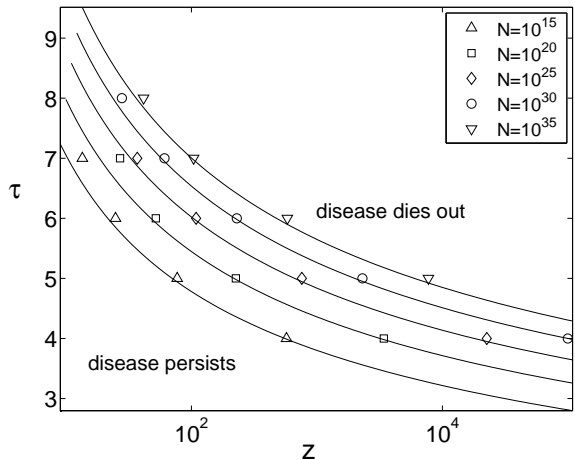}
\caption{Extinction curves for various values of system size $N$.  The
solid lines represent the asymptotic approximation for large $z$.
Here, $N$ is merely a parameter that determines the condition of
extinction for the difference equation model; it is not the number of
nodes in a simulation as in the bitstring model.  The disease is
considered extinct if the fraction infected ever falls below $1/N$.}
\label{extinct_curves}
\end{center}
\end{figure}

Figure (\ref{extinct_curves}) compares the predictions of
(\ref{extinct_predict}) with the data obtained from the map dynamics.
We see that our predictions work quite well for large $z$.  Note that
the asymptotic approximation only fits the numerical data for
unrealistically large population sizes (for smaller population sizes the
extinction boundary occurs at smaller $z$).  This is not too
troublesome, however, since our overly-simplified model cannot be
expected to quantitatively match actual population features.  Rather,
the strength of this approach is that it provides a clear picture
of a mechanism for dynamic extinction.

\section{Conclusions}

We have shown that oscillations in the number of infected
individuals in a population could be due to a mutating pathogen.  In
both of the models we have studied, oscillations occur as a consequence
of the continual introduction of novel strains, rather than the
interplay between several pre--existing strain types (for studies of
the latter, see references~\cite{lin00} and \cite{gupt98}).  We
can explain this phenomenon naturally in terms of single outbreak
epidemics: each period of oscillation can be regarded as a new epidemic
with a new strain to which few if any individuals have immunity.

In the $\tau$--dimensional map, oscillations occur only for an
intermediate range of immunity duration $\tau$.  If it is too short
(or, equivalently, if the mutation rate is too high), oscillations do
not occur because a significant pool of susceptible individuals is
always present.  Alternatively, if the duration of immunity is too
long, the infected pool oscillates with increasingly large amplitude
and ultimately becomes too small at its minimum for the pathogen to
persist.

The presence of oscillations depends on the contact rate in a similar
fashion.  For low contact rates, the susceptible pool remains large and
oscillations do not occur.  For high contact rates, large amplitude
oscillations force the number of infected individuals to such a small value that
the epidemic dies out.

We have quantified the effect of contact rate and immunity duration on
oscillatory behavior through Hopf bifurcation analysis to locate the
onset of oscillations, and with asymptotic methods to determine their
minimum value.  In both analyses, the boundary between two types of
behavior is marked by an inverse relationship between immunity
duration and contact rate.  In particular, in a population with high
contact rates, pathogen persistence requires short periods of
immunity (or high mutation rates) suggesting that highly connected
population structure could provide selective pressure for rapidly
mutating pathogens.  This could lead to diseases that are more
difficult to actively counteract through immunization programs.

There are two assumptions that could have significant impact on our
results.  First, these models operate in discrete time, where
oscillations and chaotic dynamics are known to occur more readily.
Second, we have assumed the underlying networks are fully mixed, but
some sort of quasi-static network structure may be much more
realistic.  The addition of spatial effects could dampen oscillations,
if two regions oscillated out of phase and thereby prevent global
extinction of the pathogen.

Research supported in part by the National Science Foundation,
Electric Power Research Institute, and Department of Defense.  We
thank Ken Cooke and Carlos Castillo-Chavez for helpful and interesting
discussions.


\begin{thebibliography}{99}
\bibitem{hamming} 
Hamming distance between two bitstrings is
simply the number of bits that differ between them.  For example, the
hamming distance between $0110001$ and $0100011$ is $2$.
\bibitem{footnote2} 
Because simulations are carried out on a 32 bit
computer, for simplicity $\ell=32$.  Changing this number would change
the size of the ``sequence space'' through which the strains could
move, but would not otherwise qualitatively change our results.
\bibitem{footnote3}
If $z$ is not an integer, infectives
randomly expose either $\lfloor z\rfloor$ or $\lceil z\rceil$
individuals with weighted probability such that the average number of
exposures is $z$; the model could easily be extended to other
distributions.
\end{thebibliography}
\end{document}